\documentclass[aps,prd,english,superscriptaddress,11pt,notitlepage]{revtex4}

\usepackage[utf8]{inputenc} 

\usepackage{graphicx,xcolor,overpic,mathtools}
\usepackage{amsthm,amsmath,amssymb}
\usepackage[colorlinks]{hyperref}
\definecolor{coolblack}{rgb}{0.0, 0.18, 0.39}
\hypersetup{
    colorlinks = true,
    citecolor = {coolblack},  
    urlcolor = {blue}
   }
\PassOptionsToPackage{normalem}{ulem}
\usepackage{ulem} 
\usepackage{sidenotes}
\usepackage{bm}
\usepackage{url}
\usepackage{physics}
\usepackage[makeroom]{cancel}
\usepackage{tensor}
\usepackage{mathrsfs}
\usepackage{slashed}
\usepackage{tikz}
\usepackage[mode=buildnew]{standalone}

\newcommand{\bea}{\begin{eqnarray}}
\newcommand{\eea}{\end{eqnarray}}

\newcommand{\lag}{\mathscr{L}}

\newcommand{\comment}[1]{}

\NewDocumentCommand{\evat}{sO{\bigg}mm}{%
  \IfBooleanTF{#1}
   {\mleft. #3 \mright|_{#4}}
   {#3#2|_{#4}}%
}
\usepackage{enumerate}
\usepackage{cleveref}
\usepackage{todonotes}

\usepackage{stackengine}
\stackMath

\begin{document}
\title[
]{Feshbach resonances and dynamics of BPS solitons}

\author{Alberto García Martín-Caro}
\email{alberto.garciam@ehu.eus}

\affiliation{Department of Physics, University of the Basque Country UPV/EHU, Bilbao, Spain}
\affiliation{EHU Quantum Center, UPV/EHU,}

\author{Jose Queiruga}
\email{xose.queiruga@usal.es}
\affiliation{
Department of Applied Mathematics, University of Salamanca, Spain
}

\affiliation{Institute of Fundamental Physics and Mathematics, University of Salamanca, Spain}

 \author{Andrzej Wereszczynski}
%
\affiliation{
Department of Applied Mathematics, University of Salamanca, Spain
}

\affiliation{
 Institute of Theoretical Physics,  Jagiellonian University, Lojasiewicza 11, 30-348 Krak\'{o}w, Poland
}

\affiliation{International Institute for Sustainability with Knotted Chiral Meta Matter (WPI-SKCM2), Hiroshima University, Higashi-Hiroshima, Hiroshima 739-8526, Japan}

\begin{abstract}

We demonstrate that the geodesic dynamics of BPS solitons can be modified by the excitation of Feshbach resonances, or quasi-bound modes, in a toy model of two scalar fields. A mode-generated force emerges, with a strength determined by the spectral flow of the frequency on the moduli space, and weakened by the coupling between the bound and scattering components of the resonance. Notably, spectral walls persist unaffected by the resonant mode’s exponential decay, as the decay constant vanishes at the spectral wall.
 
Our motivation comes from the 't Hooft-Polyakov monopoles, which do not present true bound states but long lived semi-bound excitations. Our findings suggest  the existence of spectral walls in the scattering of excited monopoles in three dimensions, whose trajectories may significantly deviate from the geodesic motion in the moduli space of unexcited monopoles. 

\end{abstract}
\maketitle


\section{Introduction}
BPS solitons are localized particle-like solutions parametrized by continuous parameters called {\it moduli} \cite{Manton:2004tk}. Some of them are connected with obvious space or target space transformations like translations (of the center of mass) in Poincare invariant theories or target space $O(N+1)$ rotations in nonlinear $\sigma$-models with $S^N$ target space. These are trivial moduli related to the symmetries of the theory. The nontrivial moduli do not arise from the symmetries of the action but originate from the fact that the BPS solitons obey pertinent Bogomolny equations. 

Probably, the best known example is the (2+1) dimensional Abelian-Higgs model at critical coupling. Multi-vortex solutions with topological charge $N$ can be viewed as a nonlinear superposition of $N$ unit charge vortices located at arbitrary points $z_i$, $i=1..N$ in the complex plane. These points are zeros of the Higgs field and are moduli. The position of the center of mass $z_1+...+z_N$ and a global rotation yield trivial moduli. The other are nontrivial ones. In particular, in the case of the BPS two-vortex, the nontrivial modulus is the mutual distance between the single vortices \cite{Samols:1991ne}. The BPS monopoles are another famous example of such a situation \cite{tHooft:1974kcl, Polyakov:1974ek}. 

Importantly, all configurations in the BPS sector of a classical field theory model are energetically degenerate. This means that each modulus corresponds to a zero mode. This leads to a plausible conjecture that the lowest energy dynamics corresponds to the excitation of the zero modes and is simply a transition through energetically degenerated BPS solutions. This is indeed the case. Such dynamics has found a more sophisticated formulation as geodesic motion over the moduli space, which is the space of all BPS solutions spanned by the moduli parameters \cite{Manton:1981mp}. This space is equipped with a natural metric, inherited from the standard $L^2$ norm, which amounts to force free dynamics leading to the famous $90^\circ$ scattering in head-on collisions of two BPS vortices \cite{Samols:1991ne} or BPS monopoles  \cite{Manton:1981mp}. 

For a long time, it has been believed that the geodesic flow completely explains the dynamics of BPS solitons. However, it has very recently been shown that this is not an entirely true statement. In fact, the geodesic dynamics gets significantly modified if normal massive modes hosted by solitons are excited. This happens for the BPS vortices where, in the case of two-vortex solution with the lowest mode excited, the $90^\circ$ head-on scattering is replaced by multi-bounce windows forming a chaotic (probably fractal) pattern \cite{Krusch:2024vuy, AlonsoIzquierdo:CollectiveVortices}, very similar to the fractal structure in the kink-antikink collisions in $\phi^4$ theory \cite{Campbell:1983xu, Sugiyama:1979mi, Manton:2021ipk}. The actual final state depends on the amplitude $A$ of the excited mode. This phenomenon is explained by a mode generated force which appears due to the new potential energy component $E_{mode}=\frac{1}{2}\omega^2(d)A^2$ \cite{AlonsoIzquierdo:CollectiveVortices}, where the frequency of the mode {\it depends} on the distance $d$ between the two vortices \cite{Alonso-Izquierdo:2023cua}. This is because of the fact that, although energetically degenerated, the solutions in the same BPS sector do not necessarily share the same spectrum of linear perturbations. In general, the spectrum of the differential operator $\mathcal{L}(q)$ associated to linear perturbations on the background of an element on the moduli space will change smoothly with the moduli coordinates $q$. The smooth flow of the spectrum of a linear fluctuation operator associated to the flow on the moduli space is known as \emph{spectral flow}. The derivative of such frequencies with respect to the moduli space coordinates determines the strength and the sign of the ``force" felt by the solitons. Such effect has been reported before for the scattering of wobbling kinks \cite{Adam:2019xuc} and recently for excited vortices \cite{AlonsoIzquierdo:CollectiveVortices}. 

In some critical cases, the mode can cease to exist. This occurs if the mode enters the continuum at some point in moduli space. In such cases, solitons typically encounter a localized barrier, a phenomenon known as \emph{spectral wall}, which was first discovered in kink-antikink collisions \cite{Adam:2019xuc} and has only recently been studied for scalar fluctuations of BPS vortices \cite{Alonso-Izquierdo:SWvortices}. For a critical value of the amplitude of the mode, the solitonic system (e.g., the BPS 2-vortex) forms a long-living quasi-stationary state with solitons located exactly at the distance corresponding to the point on the moduli space, where the mode hits the mass threshold.

To conclude, it is nowadays clear that dynamics of the BPS solitons (e.g., kinks or vortices) is a combined effect of a force free flow (associated to geodesics of the moduli space metric) and the mode-generated forces due to the spectral flow of eigenfrequencies of excitations over the moduli space with additional contribution from a possible spectral wall. 

However, not all BPS solitons possess massive normal modes. For example, the 't Hooft-Polyakov monopoles have infinitely many quasi-normal modes (QNM) representing (linearly unstable) long-living resonances \cite{Forgacs:2003yh,Russell:2010xx}. These particular QNM are Feshbach resonances \cite{Feshbach}, which appear generically in multi-channel scattering problems involving the coupling between fields with different mass thresholds  \cite{PhysRev.147.73}. In the context of topological solitons, they also exist as the excitation of Abelian-Higgs vortices outside of the BPS regime \cite{Alonso-Izquierdo:2024tjc}. 

Feshbach resonances are a particular kind of a broad class of long-lived resonances, or \emph{quasi-normal modes}, which appear typically in field theory as solutions of eigenvalue problems with complex eigenvalues, representing dissipative processes.  
These solutions are characteristic of linearized perturbation equations around localized objects such as solitons, but also black holes and branes \cite{Berti:2009kk}. The effects of quasi-normal modes in the dynamics of solitons, such as kink-antikink collisions, have also been reported previously in \cite{Dorey:2017dsn, Campos:2019vzf}.

The aim of the present work is to understand the impact of the Feshbach resonances on the standard geodesic motion of BPS solitons. We will investigate this problem using a (1+1) dimensional model with two scalar fields. This significantly simplifies numerical computations. Nevertheless, it may allow us to get some intuition about the BPS monopoles. 

The structure of the paper is the following. In \cref{sec:model} we present the model, the solutions of the corresponding BPS equations and the moduli space metric. In \cref{sec:spectral_flow} we compute the effective potential for linear perturbations, and discuss the spectral flow of the frequencies associated to the eigenmodes of the uncoupled system. In \cref{sec:gamma} we couple the two channels of the linear perturbation problem and compute the flow of decay constant on the moduli space. We then numerically solve the time evolution of excited BPS configurations in \cref{sec:numerical} and explain the observed dynamics using the flow of the frequencies and the decay constant of the Feshbach resonances. In \cref{sec:conc} we summarize our findings.
\section{The model}
\label{sec:model}

We consider the following model, which consists of two coupled real scalar fields with Lagrangian:
\begin{equation}
    \lag = \frac{1}{2}[\partial_\mu\phi\partial^\mu\phi+\partial_\mu\chi\partial^\mu\chi]-V(\phi,\chi),
\label{modelLag}
\end{equation}
where 
\begin{equation}
    V(\phi,\chi)=\frac{1}{2} \left(1-\phi ^2-r \chi ^2\right)^2+2 r^2 \chi ^2 \phi ^2
\end{equation}
is a potential and $r\in \mathbb{R}$ is a coupling constant.

The model \eqref{modelLag} was first considered by Bazeia et al. \cite{Bazeia:1995en,Bazeia:1996np} and its BPS sector was further explored by Shifman and Voloshin \cite{ShifmanVoloshin} and Alonso-Izquierdo et al. \cite{Alonso-Izquierdo:2013isa}.

The potential can be rewritten as
\begin{equation}
    V(\phi,\chi)=\frac{1}{2}\qty[(\partial_\phi W)^2+(\partial_\chi W)^2]
\end{equation}
in terms of the superpotential
\begin{equation}
    W(\phi,\chi)=-r \chi ^2 \phi -\frac{\phi ^3}{3}+\phi,
\end{equation}
so that the static energy functional becomes
\begin{equation}
        E = \int_{-\infty}^\infty \left( \frac{1}{2}[(\phi'-\partial_\phi W)^2+(\chi'-\partial_\chi W)^2]+dW \right)dx,
\end{equation}
and the minimal energy configurations are thus attained by solutions of the system of the two coupled first order BPS (Bogomolny) equations:
\begin{align}
    \phi'=\partial_\phi W=1-\phi^2-r\chi^2,\qquad
    \chi'=\partial_\chi W=-2r\phi\chi.
\label{autn}
\end{align}
The solutions in the BPS sector are energetically degenerate as long as the topology (boundary conditions) are kept fixed.  The energy reads $E=W(\phi(\infty),\chi(\infty))-W(\phi(-\infty),\chi(-\infty))$.
As described in \cite{Alonso-Izquierdo:2013isa}, the model presents a plethora of solutions of very different nature depending on the value of the coupling $r$. We will be interested in the particular case where  $r=1/4$, for which the most general non-trivial solutions involve a kink in one component and a non-topological lump in the other. Following \cite{Alonso-Izquierdo:2013isa}, we integrate the
equation of the orbits for the autonomous system \eqref{autn}. Furthermore, imposing the appropriate boundary conditions the following exact solutions can be found:
\begin{eqnarray}
    \phi(x; a,\gamma)&=&\frac{\sinh (x-a)}{\cosh(x-a)+\gamma^2},\\
    \chi (x; a,\gamma )&=&\frac{2\gamma}{\sqrt{\gamma^2+\cosh(x-a)}}.
\end{eqnarray}
There are two moduli parameters of the BPS sector $(a,\gamma) \in \mathbb{R}^2$. The first one, $a$, denotes the position of the center of mass of the composed soliton form by these two fields. This is a trivial modulus arising due to the translation invariance of the Lagrangian. The second modulus $\gamma$ non-trivially deforms the profile. 

\begin{figure}
    \includegraphics[scale=0.4]{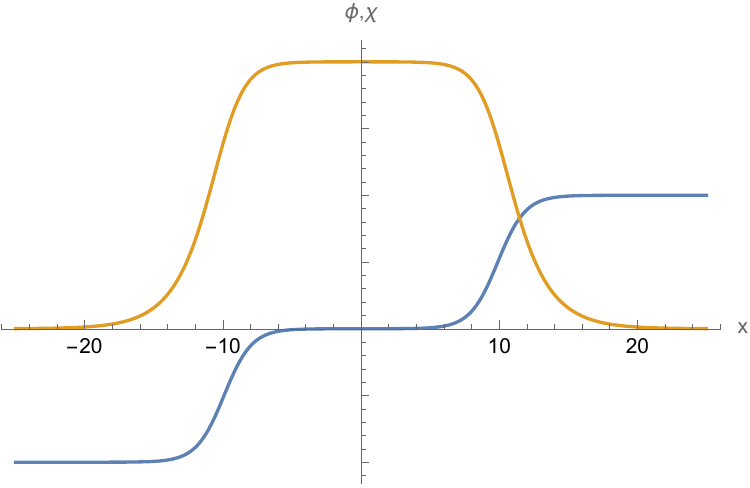}
    \includegraphics[scale=0.4]{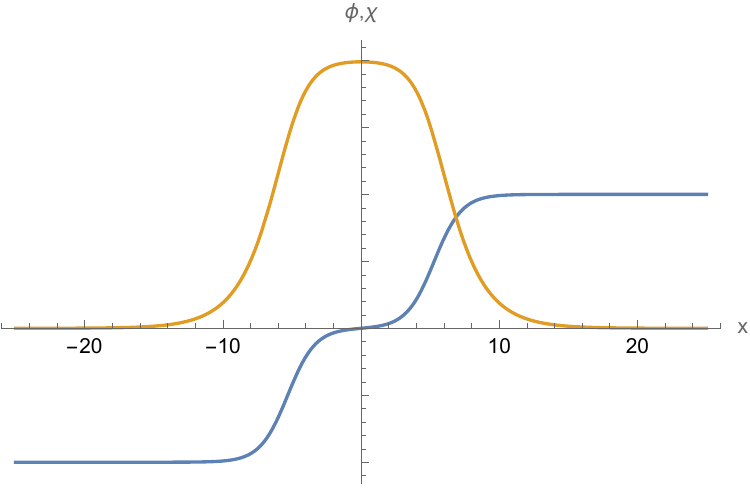}
    \includegraphics[scale=0.4]{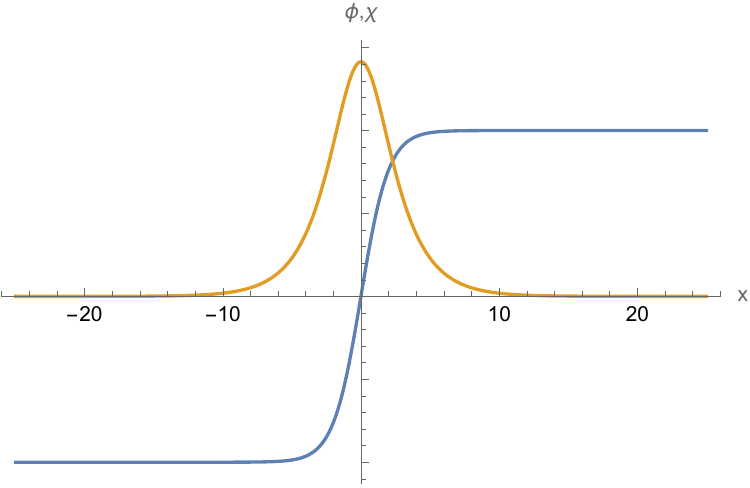}
     \includegraphics[scale=0.4]{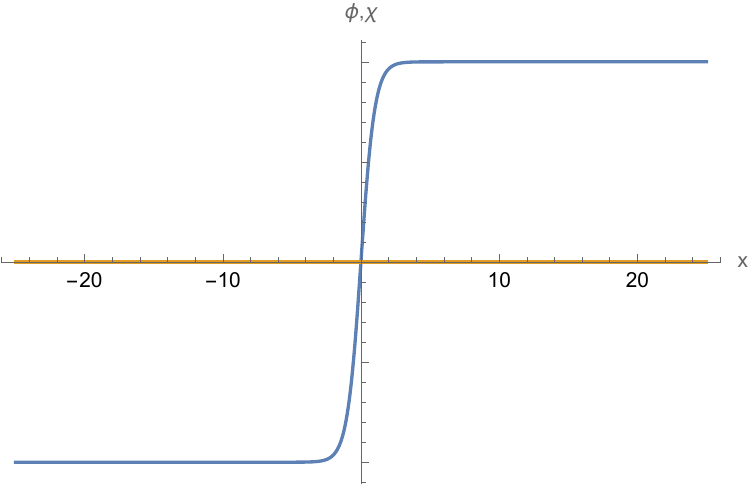}
    \includegraphics[scale=0.4]{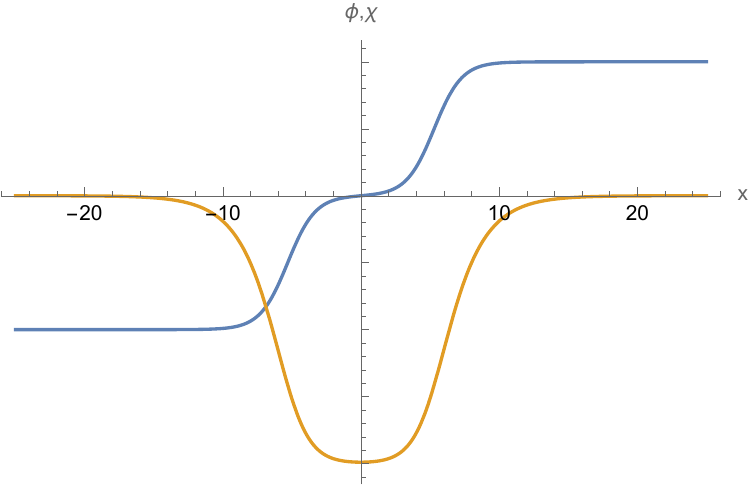}
    \includegraphics[scale=0.4]{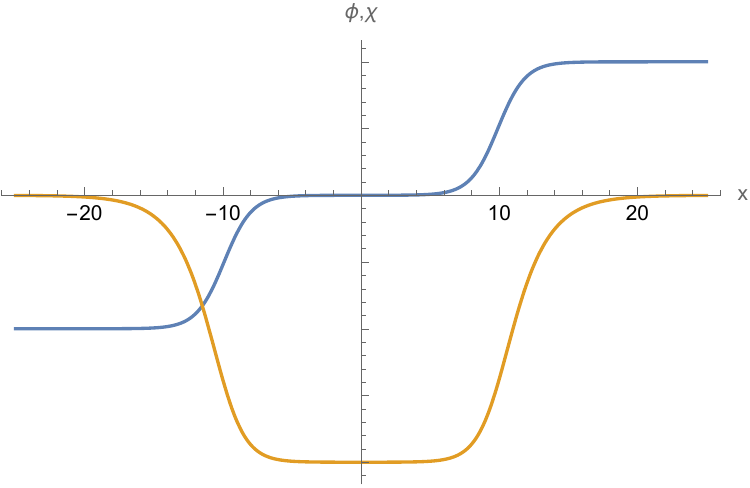}
    \caption{BPS solutions for $\gamma=100, 10, 1, 0, -10, -100$ and $a=0$. Blue and orange curves are $\phi$ and $\chi$ field respectively. }
    \label{fig:Profiles}
\end{figure}

For $\gamma \to \infty$ the solution decomposes into two infinitely separated semi-kinks in $\phi$ field, $\Phi_{(-1,0)} + \Phi_{(0,1)}$, and semi kink-antikink in $\chi$ field, $X_{(0,2)}+X_{(2,0)}$, see Fig. \ref{fig:Profiles}. Here the subscript denotes the asymptotic value of the field at infinities. Concretely they are
\begin{eqnarray}
   \Phi_{(-1,0)} &=&-\frac{1}{1+e^{x-x_1}}, \;\;\; \Phi_{(0,1)}=\frac{1}{1+e^{-(x-x_2)}} 
   \label{eq:asympt}
   \\
  X_{(0,2)} &=& \frac{2}{\sqrt{1+e^{-(x-x_1)}} }, \;\;\; X_{(2,0)} = \frac{2}{\sqrt{1+e^{x-x_2}} }
\end{eqnarray}
where $x_{1,2}= \mp \ln (2\gamma^2)$ are the position of the half-solitons.

As $\gamma$ decreases the constituent semi-(anti)kinks approach each other forming the on-top configuration for $\gamma=0$. Here, we find the usual $\tanh(x)$ kink in $\phi$ field and uniform $\chi\equiv 0$. Then, as $\gamma$ further decreases to negative values, the semi-(anti)kink separate once again. Now, the BPS solution in $\chi$ field is semi antikink-kink i.e., it is a negative lump. Thus, asymptotically, for $\gamma= \infty$ we get a two different solitonic bound states (molecules). The first one consists of half-kinks in $\phi$ and $\chi$ fields (located on the l.h.s). The second one consists of a half-kink in $\phi$ and a half-antikink in the $\chi$ field (located on the r.h.s). For $\gamma=-\infty$ we find the half-kink in $\phi$ and half-antikink in $\chi$ (located on the l.h.s) and half-kinks in the $\phi$ and $\chi$ fields (located on the r.h.s). These ``molecules" may be viewed as equivalents of the BPS multivortex with unit charge. 

If we take $\gamma \gg 1$ as an initial state and excite the nontrivial zero mode, i.e. boost the half-kinks towards each other, they will scatter by passing through the BPS solutions corresponding with decreasing of $\gamma$. The actual dynamics as obtained in field theory simulations is very well captured in terms of the force free geodesic motion.

To see this at a quantitative level we follow the usual prescription and construct a collective coordinate model (CCM). Thus,  we insert the static BPS solution into the original field theory and perform the spatial integration with the assumption that the moduli are time-dependent coordinates. This leads to the following effective Lagrangian
\begin{equation}
    L[a,\gamma]= \frac{1}{2}g_{aa}\dot{a}^2 + \frac{1}{2} g_{\gamma \gamma }\dot{\gamma}^2 - M,
\end{equation}
where the moduli space metric components are
\begin{equation}
g_{aa}\equiv M = \frac{4}{3},  
\end{equation}
and
\begin{equation}
g_{\gamma \gamma}  =-\frac{8}{3 \left(1-\gamma^4 \right)^{5/2}}\left(6\arctan \left(\frac{\gamma^2-1}{\sqrt{1-\gamma^4}}\right)+\gamma^2\left(5-2\gamma^4\right)\sqrt{1-\gamma^4}\right).
\end{equation}
The off diagonal component vanish due to the orthogonality of the zero modes associated with the moduli. Additionally, $M=\frac{4}{3}$ is the mass of the BPS soliton. 

Not surprisingly, we found an excellent agreement between the CCM dynamics and  the numerical integration of the Euler-Lagrange equation of the field theory with the solution of the CCM. The initial state is the BPS solution with large $\gamma(t=0)\equiv\gamma_0$, describing ``molecules" consisting of two well separated half-kinks, which are boosted towards each other with initial velocity $v$. Without loosing generality, we consider only collisions with center of mass at the origin, $a=0$. Again, this can be treated as a $(1+1)$ dimensional counterpart of the geodesic motion of BPS vortices.

\section{Spectral flow over moduli space}
\label{sec:spectral_flow}
We now consider linear perturbations on top of the static BPS solutions $\Phi(x; a,\gamma)=(\phi(x;  a,\gamma),\chi(x; a,\gamma) )$, i.e. the perturbed field
\begin{equation}
  \Psi(x; a,\gamma)= \Phi(x; a,\gamma)+\epsilon e^{i\omega t}\delta\Phi(x; a,\gamma) 
\end{equation}
must be a solution of the equations of motion to first order in $\epsilon$, which is equivalent to imposing that $\delta\Phi(x; a,\gamma) =
(\eta(x; a,\gamma),\xi(x; a,\gamma))$ must be an eigenstate of the associated Schrödinger-like differential operator:
\begin{equation}
    \mathcal{L}_{a,\gamma}\delta\Phi=\omega^2(a,\gamma)\delta\Phi,
\end{equation}
where $\omega \in \mathbb{R}$ is a real frequency.
Due to the translational invariance of the theory, the spectral problem does not depend on the position of the center of mass $a$. Thus, we can safely assume $a=0$. Then, 
\begin{equation}
   ( \mathcal{L}_{\gamma})_{ij}=-\delta_{ij}\dv{^2}{x^2}+V_{ij}[\phi(x; 0,\gamma)],
   \label{Operator}
\end{equation}
and
\begin{align}
   V_{\eta\eta}[\phi(x;\gamma)]&= \frac{\gamma ^4-\gamma ^2 \cosh (x)+2 \cosh (2 x)-4}{\left[\gamma ^2+\cosh (x)\right]^2},\label{Veta2}\\[2mm]
   V_{\xi\xi}[\phi(x;\gamma)]&= \frac{3 \left(\gamma ^4-1\right)}{4 \left[\gamma ^2+\cosh (x)\right]^2}+\frac{1}{4},\label{Vetaxi}\\[2mm]
     V_{\eta\xi}[\phi(x;\gamma)]&=
    \frac{3 \gamma  \sinh (x)}{\left[\gamma ^2+\cosh (x)\right]^{3/2}}.\label{Vxi2}
\end{align}

\begin{figure}[htb!]
    \centering
\includegraphics[scale=0.7]{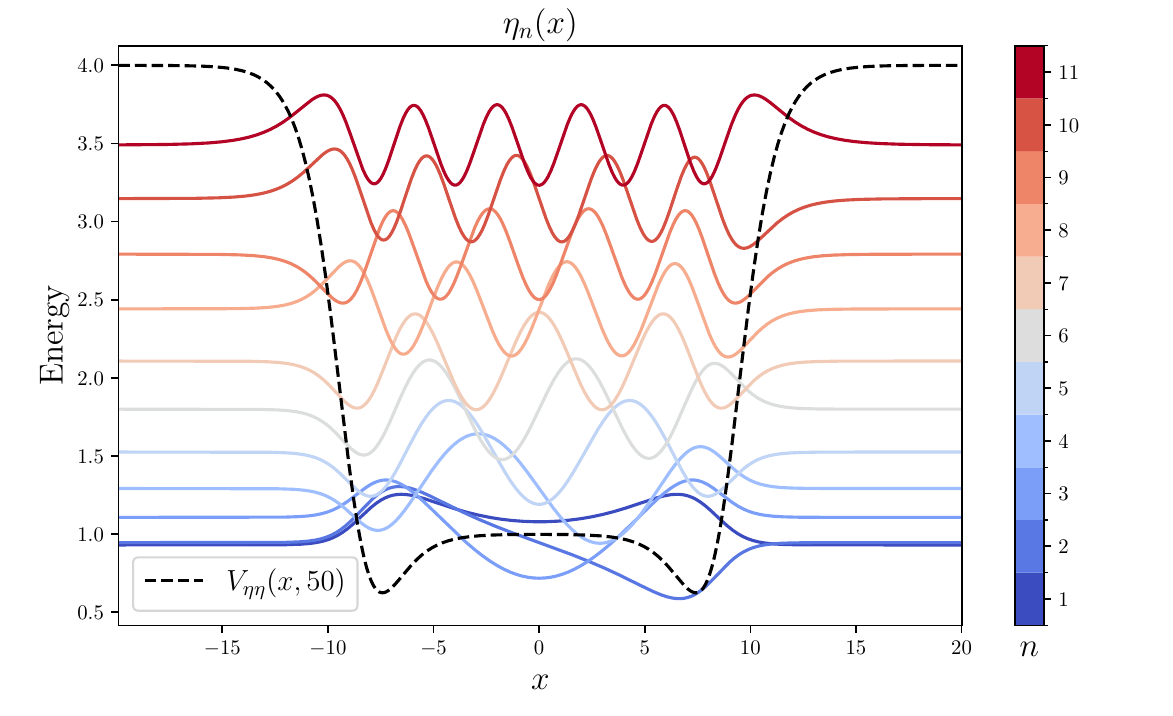}
    \caption{Bound modes $\eta_n$ in the truncated spectral problem for $\gamma=50$.}
    \label{fig:Sp_mode}
\end{figure}

First of all, we observe that the differential operator \eqref{Operator} is that of a coupled, two-channel scattering problem for the fields $(\eta,\xi)$ that reduces to a pair of decoupled, free wave operators at long distances, as can be deduced by taking the limit
\begin{equation}
    \lim\limits_{x\to\infty}V_{ij}[\phi(x;0,\gamma)]=\mqty(4&0\\0&1/4)\equiv \mqty(m_\eta^2&0\\0&m_\xi^2).
\end{equation}
Importantly, the two mass thresholds are different and a bound state in one channel, here in the $\eta$ perturbation, can be a scattering state in the second channel,that is, in the $\chi$ direction. This is the set-up in which we can find Feshbach resonances. To approximate such a resonant state we consider a decoupled limit of the spectral problem and find only bound eigenstates in the $\mathcal{L}_{\eta \eta}$ channel. In Fig. \ref{fig:Sp_mode} we show the bound modes $\eta(x)$ in such a truncated linear problem hosted by the BPS solution with $\gamma=50$. Note that all of them lay {\it above} the mass threshold of the $\chi$ field channel. 

\begin{figure}[htb!]
\hspace*{-1cm}\includegraphics[scale=0.5]{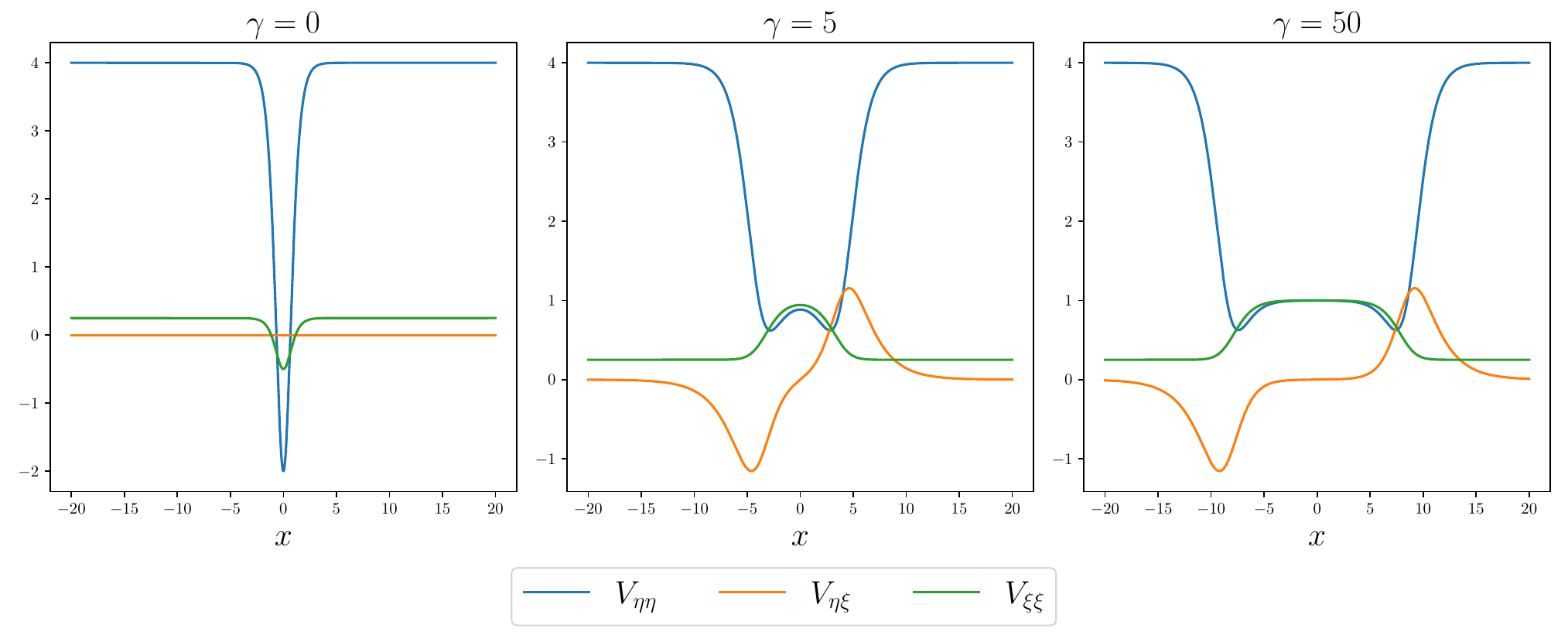}
    \caption{Matrix elements of the potential for linearized perturbations at $\gamma=\left\{0, 5, 50\right\}$.}
    \label{fig:Pot}
\end{figure}
\begin{figure}
    \centering
   \includegraphics[scale=0.75]{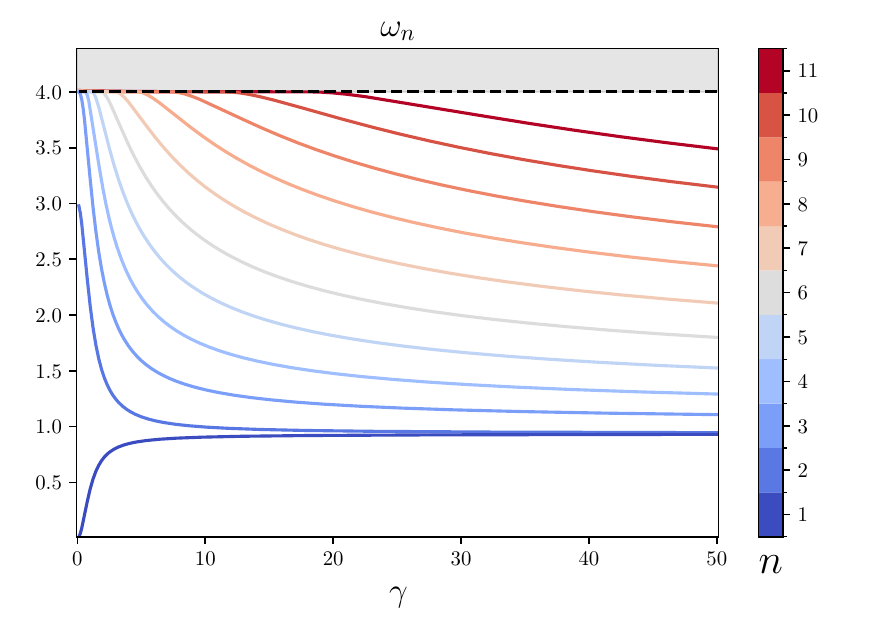}
    \caption{Spectral flow for the  first eleven eigenvalues of the truncated operator $(\mathcal{L}_\gamma)_{\eta\eta}$. }
    \label{fig:Sp_flow}
\end{figure}

From the point of view of dynamics triggered  by the Feshbach resonances, the most important property is the change of the potential well arising in the $V_{\eta \eta}$ part of the spectral problem as we change the modulus $\gamma$, see Fig. \ref{fig:Pot}. Indeed, $V_{\eta \eta}$ has the shape of a potential well whose width increases arbitrary with growing $|\gamma|$. Therefore, it can support more and more modes as $\gamma$ grows, see Fig. \ref{fig:Sp_flow}. As $\gamma$ decreases to 0, the potential well closes up and, one after the other, the modes enter the continuum.

The meaning of that is the following. When the BPS solution consists of a molecule of two largely separated half kinks, it can host a large number of Feshbach resonances. This number goes to infinity as $|\gamma|\to \infty$ i.e., as the distance between the sub-kinks grows. The frequencies of the Feshbach resonances also change as we move in the moduli space, $\omega^2=\omega^2(\gamma)$. Except the first mode, they decrease with $|\gamma|$. Eventually, each of the modes with $n \ge 3$ hits the mass threshold at some value of $\gamma$, where they are no longer a bound mode in the $\eta$ channel.

Qualitatively similar flow of the spectral structure was previously observed in various BPS field theories in one and two spatial dimensions, however, in the more standard case of normal modes. For example, a nontrivial dependence of the spectral structure on the position on the moduli space was observed in the Abelian Higgs model at the critical coupling \cite{Alonso-Izquierdo:2023cua}. The only difference is that the normal modes are now replaced by the Feshbach resonances. One may also notice a particular similarity with flow of the modes in the non-BPS kink-antikink moduli space in the $\phi^6$ model, where, due to the different values of the mass of the vacuum excitations, a similar potential well forms between the solitons 
\cite{phi6scattering}. These similarities lead to two conjectures.

First of all, one may expect that the {\it mode generated force} will appear. This comes from the fact that at the leading quadratic order a (normalized) bound mode contributes to the potential energy as \cite{AlonsoIzquierdo:CollectiveVortices}
\begin{equation}
V_{int} = \frac{1}{2} \omega^2(\gamma) A^2,    
\end{equation} 
where $A$ is the amplitude of the mode. 
Since for $n\geq 3$ the frequency grows as $|\gamma|$ decreases the subkins forming solitonic molecules can feel a mutually repulsive force. For a sufficiently large amplitude of the mode, this force may overcome the kinetic motion and the solitons in a molecule may be back-scattered. 

Secondly, a {\it spectral wall} may appear. This is a barrier in the motion of a soliton due to the transition of a mode from the discrete to the continuum spectrum \cite{Adam:2019xuc}. If the corresponding mode is excited then for a critical value of its amplitude, the soliton forms a long living quasi-stationary state whose spatial position exactly agrees with the position on the moduli space where the mode enters the continuum spectrum. 

However, for these two conclusions to hold in our present set up one has to take into account the fact that we are dealing with Feshbach resonances, not proper normal modes. The Feshbach resonances can be thought of as being composed by a bound state in one channel and a scattering mode in the other. These two components are coupled already at linear order, which leads to a decay of the bounded component in an exponential manner
\begin{equation}
A=e^{-\Gamma t}A_0,     
\end{equation}
where $A_0$ is the initial amplitude and $\Gamma$ is the decay constant. Note that the usual normal modes also decay eventually, but this is due to higher order self interactions, resulting in a much slower, power-like decay \cite{Manton:1996ex}.

Of course, the value of the decay constant is of fundamental importance for the dynamics of the excited soliton. If $\Gamma$ is large, the mode may decay so rapidly that the whole energy stored in the resonance is almost instantaneously radiated out. Effectively, we would rather arrive at a system consisting of a soliton immersed in radiation background. If $\Gamma$ is relatively small the excited soliton may have time to experience the mode-generated force and even the spectral wall.

Furthermore, also the decay constant $\Gamma$ can flow over moduli space, which may have a non-trivial impact on the dynamics. This will be investigated in the next section.

\section{Flow of the decay constant} 
\label{sec:gamma}
It is an obvious but very interesting observation that not only the (real) frequency of the Feshbach resonance varies with the position on the moduli space but the same happens with the decay constant $\Gamma$, which also depends on $\gamma$. 

Consider the system of linearized equations defined by \cref{Operator}, but in the decoupling limit, i.e. $V_{\xi\eta}=V_{\eta\xi}=0$. Then, the system reduces to a set of two independent equations, one for each of the perturbation fields. As we have seen, the truncated subsystem for the perturbation $\eta$ presents a discrete set of bounded eigenstates in its spectrum, i.e.
\begin{equation}
    \eta_n''(x)-\qty(\frac{\gamma ^4-\gamma ^2 \cosh (x)+2 \cosh (2 x)-4}{\left[\gamma ^2+\cosh (x)\right]^2}-\omega_n^2 )\eta_n(x)=0,
\end{equation}
whereas the subsystem associated to the perturbation in the other channel given by 
\begin{equation}
    \ddot{\xi}(x,t)-\xi''(x,t)+\qty(\frac{3 \left(\gamma ^4-1\right)}{4 \left[\gamma ^2+\cosh (x)\right]^2}+\frac{1}{4})\xi(x,t)=0,
\label{eq:xidott}
\end{equation}
only presents scattering states of the form \begin{equation}
    \xi(x,t)=f_q(x)e^{i(qx-\omega t)},
    \label{eq:shapexi}
\end{equation} where the spatial profile $f_q(x)$ satisfies the following equation:
\begin{equation}
    -f''_q(x)+2iq f'_q(x)+\qty(\frac{3 \left(\gamma ^4-1\right)}{4 \left[\gamma ^2+\cosh (x)\right]^2})f_q(x)=0,
    \label{eq:efe_n}
\end{equation}
with $q=\sqrt{\omega^2-1/4}$.

Therefore, if we slowly switch up the coupling part of the potential, the bound state $\eta_n$ will start loosing its energy to the $\xi$ channel and will be radiated to infinity. Then, the mode becomes a quasi-normal mode.

Following the same procedure as in \cite{Manton:1996ex,Forgacs:2003yh}, one can find approximately the solutions in the coupled problem by successive iterations, starting from the solutions to the uncoupled system. In particular, the back-reaction on the bound state $\eta_n$ due to energy leaking through the other channel can be also computed by considering a source term in \eqref{eq:xidott} of the form:
\begin{equation}
    \ddot{\xi}(x,t)-\xi''(x,t)+\qty(\frac{3 \left(\gamma ^4-1\right)}{4 \left[\gamma ^2+\cosh (x)\right]^2}+\frac{1}{4})\xi(x,t)=-\frac{3 \gamma  \sinh (x)}{\left[\gamma ^2+\cosh (x)\right]^{3/2}}A_n\eta_n(x)e^{-i\omega_n t},
    \label{eq:xi_source}
\end{equation}
namely, the source term is constructed from the $n-th$ bound state solution in the decoupled limit, excited with constant amplitude $A_n$, and coupled through the non-diagonal part of the potential \eqref{Vxi2}. 

We now look for a solution for $\xi$ that shares the same time dependence of the source, hence, of the form \eqref{eq:shapexi} with frequency $\omega_n$. On the other hand, if we (numerically) compute the solutions of the homogeneous part, or equivalently, the solutions of \cref{eq:efe_n}, we can write the relevant solution of \eqref{eq:xi_source} as
\begin{equation}
    \xi(x,t)=e^{-i\omega_nt}A_n\int\limits_{-\infty}^\infty G(u,x) V_{\xi\eta}[\Phi(u,\gamma)]\eta_n(u)du
\end{equation}
where $G(u,x)$ is the Green's function of the homogeneous problem \eqref{eq:xidott}:
\begin{equation}
    G(u,x)=-\frac{1}{W}\qty[f_{-q_n}(u)f_{q_n}(x)e^{-iq_n(u-x)}\theta(u-x)-f_{-q_n}(x)f_{q_n}(u)e^{-iq_n(x-u)}\theta(x-u)],
\end{equation}
and 
\begin{equation}
    W=-2iq_n f_{q_n}(x)f_{-q_n}(x)+f_{q_n}(x)f'_{-q_n}(x)-f'_{q_n}(x)f_{-q_n}(x)
\end{equation}
is the Wronskian. Since the Wronskian of a Sturm-Liouville problem without first derivatives must be a constant, we can evaluate it at spatial infinity:
\begin{equation}
    W=-2iq_n |f_{q_n}(\infty)|^2 .
\end{equation}

Now, we will be interested in the asymptotic form of such solution at infinity. Integrating the conservation of the energy momentum tensor $\partial_\mu T^\mu_\nu=0$, we deduce that the variation of the total energy stored in the bound state is given by the flux of radiation at infinity:
\begin{equation}
    \dv{}{t}\qty(\int_{-\infty}^\infty T^0_0 dx)=-\lim\limits_{x\to \infty}T^x_0+\lim\limits_{x\to -\infty}T^x_0
\label{energybalance}
\end{equation}
Therefore, asymptotically we have
\begin{equation}
    \lim\limits_{x\to\infty}\xi(x,t)\approx e^{-i(\omega_n t+q_nx)}\frac{iA_n}{2q_n|f_{q_n}(\infty)|}\int\limits_{-\infty}^\infty f_{q_n}(u) e^{iq_nu}V_{\xi\eta}[\Phi(u,\gamma)]\eta_n(u)du
\end{equation}
and thus 
\begin{equation}
    \lim\limits_{x\to \pm \infty}T^x_0=\pm c_n A_n^2
\end{equation}
where
\begin{equation}
    c_n=\frac{\omega_n }{8\sqrt{\omega_n^2-1/4}|f_{q_n}(\infty)|^2}\abs{\int_{-\infty}^\infty f_{q_n}(u) e^{iq_nu}V_{\xi\eta}[\Phi(u,\gamma)]\eta_n(u)du}^2.
    \label{decayconst}
\end{equation}

On the other hand, the relevant part of the total energy in the left hand side of \eqref{energybalance} is just the energy of the bound state, which, after averaging over one oscillation, is simply given by $E=\tfrac{1}{2}\omega_nA_n^2$. Then, if we promote this amplitude to a time dependent one, the equation for the energy balance becomes
\begin{equation}
    \dot{A}=-\Gamma_n A_n, \qquad \Gamma_n\equiv \frac{2 c_n}{\omega_n}.
\end{equation}
reproducing the exponential decay law. 

\begin{figure}
\hspace*{-1.8cm}\includegraphics[width=0.55\linewidth]{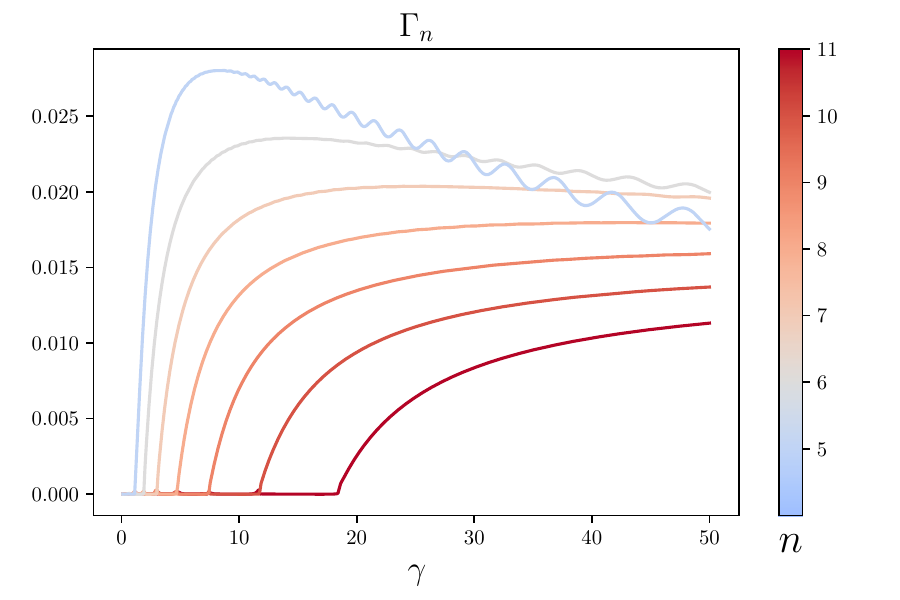}
\includegraphics[width=0.55\linewidth]{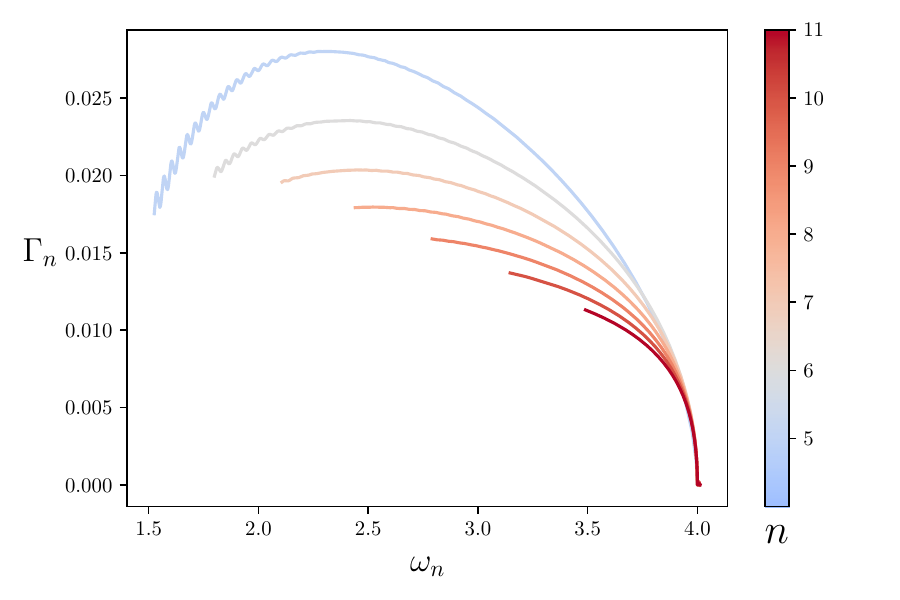}
    \caption{Left: flow of the  bound state decay constant with the moduli space coordinate for several modes. Right: decay constant of each resonance as a function of the eigenfrequency}
    \label{fig:Gamma_flow}
\end{figure}

In Fig. \ref{fig:Gamma_flow}, left panel, we show how the decay constant $\Gamma$ depends on the moduli coordinate $\gamma$ for the modes with $n\geq 4$. The following remarks are in order.

First of all, we observe that the decay constants have rather small values. Thus, the modes do not lose their energy too rapidly. Hence, the effects known from solitons with normal modes have a chance to be observed here. This behaviour should be further enhanced if we consider a higher mode (with bigger $n$). Indeed, a higher mode (at least for $n\geq 5$ and in the range $|\gamma| \leq 50$) possesses smaller $\Gamma$. 

A somewhat striking result is that all $\Gamma$'s vanish as the bound part of the mode approaches the corresponding mass threshold. A similar behavior has been previously observed in the excitation of vortices \cite{Blanco-Pillado:2021jad, Alonso-Izquierdo:2024tjc}. It means that in the vicinity of this point, which is exactly the expected location of the corresponding spectral wall, the mode becomes more and more stable with $\Gamma$ suppressed to zero. In other words, {\it the spectral wall acts as a stabilizer of the resonances}. This suggests that the spectral wall phenomenon may exist also in the case of the Feshbach resonances and simultaneously explains why the quasi-stationary solution formed on the spectral wall may live for a very long time.

A qualitative explanation for this phenomenon is as follows: as the mode $\eta_n$ approaches the mass threshold it spreads out while maintaining its normalization. On the other hand, the potential $V_{\eta\xi}$ is proportional to the spatial derivative of the $\chi$-field which has a fixed support around the soliton. Therefore, close to the mass threshold, the overlap between $\eta_n$ and $V_{\eta\xi}$, given by the integral \eqref{decayconst}, which is proportional to $\Gamma_n$, tends to zero. This is a generic mechanism which  should apply to other BPS solitonic models with Feshbach resonances. Importantly, we expect that it also applies to the BPS monopoles.

In fact, a similar behaviour of the decay constant was also observed in the case of the resonant modes of the unit charge BPS monopole. The resonances with the real frequency approaching the mass threshold of the vector channel have the decay constant decreasing to zero \cite{Forgacs:2003yh}. This shows that our toy model is quite well tailored to qualitatively study the dynamics of the BPS monopoles.

In Fig. \ref{fig:Gamma_flow}, right panel, we show how the relation between the frequency and the decay constant of the mode as we flow on the moduli space i.e., as the half-kinks of a molecule approach each other. 

\section{Exciting Feshbach resonances} 
\label{sec:numerical}
Now we will verify our analysis and investigate how the BPS solitons with the excited Feshbach resonances actually interact. 

The equations of motion derived from the Lagrangian \eqref{modelLag} are given by
\bea
    \partial_\mu\partial^\mu \phi - 2\phi (1-\phi^2-\frac{1}{4}\chi^2) + \frac{1}{4}\chi^2 \phi &=& 0, \\
    \partial_\mu\partial^\mu \chi - \frac{1}{2}\chi (1-\phi^2-\frac{1}{4}\chi^2) + \frac{1}{4}\chi \phi^2 &=& 0. 
\eea
We will study the dynamics of excited BPS configurations by preparing initial conditions of the form:
\begin{align}
    \Psi(x,t=0)&=\left(
    \begin{array}{c}
         \phi(x;\gamma_0)  \\
         \chi(x;\gamma_0) 
         \end{array}
         \right)+A_0\left(
          \begin{array}{c}
         \eta_n(x,\gamma_0) \\
        0 
         \end{array}
         \right),
\label{init_conds1}
\\[2mm]
\dot\Psi(x,t=0)&=\left(
    \begin{array}{c}
         \partial_\gamma\phi(x;\gamma)  \\
        \partial_\gamma \chi(x;\gamma) 
         \end{array}
         \right)\Bigg\rvert_{\gamma_0} \dot{\gamma_0}+A_0\left(
          \begin{array}{c}
         \partial_\gamma\eta_n(x,\gamma) \\
        0 
         \end{array}
         \right)\Bigg\rvert_{\gamma_0} \dot{\gamma_0}, 
\label{init_conds2}
\end{align}
where $\eta_n(x,\gamma_0)$ is the $n-$th bound state associated to the truncated operator $(\mathcal{L}_{\gamma_0})_{\eta\eta}$.  Moreover, $\dot\gamma_0$ is the initial velocity in the moduli space, which is related to the real space velocity of the sub-kinks through
\begin{equation}
    v_0=\frac{-\gamma_0+\sqrt{16 \dot\gamma_0^2+\gamma_0^2}}{4 \dot\gamma_0}.
    \label{v0_rel}
\end{equation}
The above relation can be obtained by taking into account the Lorentz boost factor in the asymptotic expression. In the non-relativistic limit ($\dot\gamma_0\ll\gamma_0$) , \cref{v0_rel} yields $
    v_0=2\frac{\dot{\gamma}_0}{\gamma_0}$. We will assume that \cref{init_conds1,init_conds2} constitute a good initial approximation to describe the excitation of a Feshbach resonance of the system.

\begin{figure}[htb!]
        \includegraphics[width=0.7\textwidth]{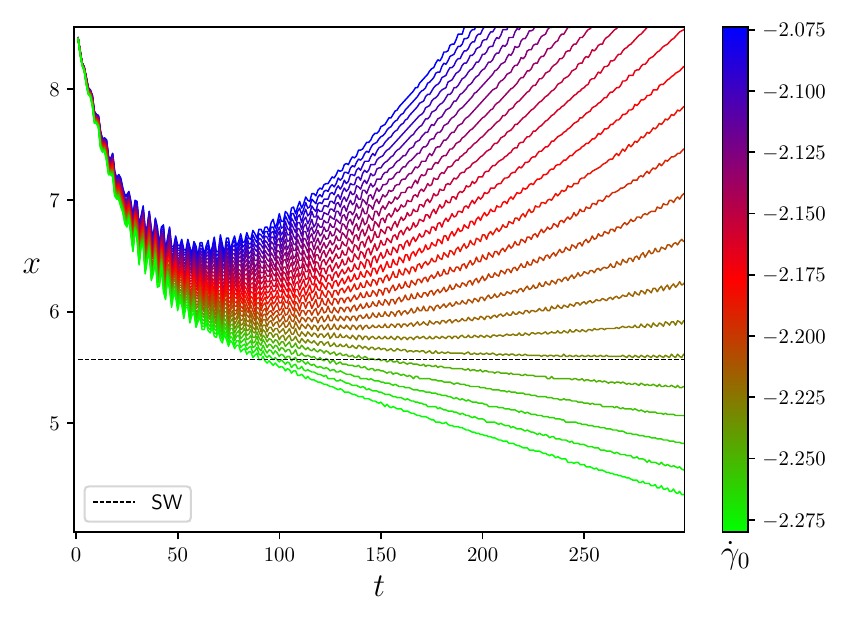}
         \includegraphics[width=0.7\textwidth]{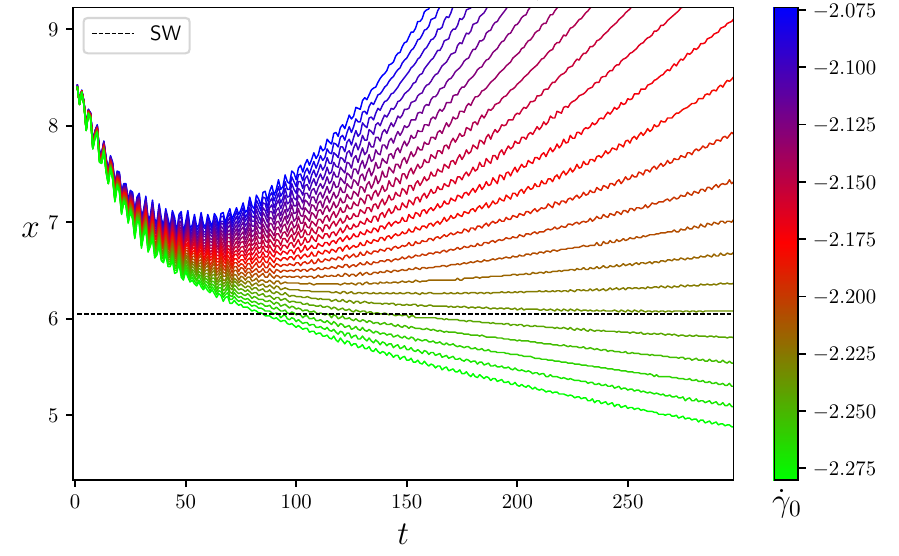}
    \caption{Trajectories of a semi-kink which forms the BPS solution with the excited mode: $\eta_{10}$ (upper) and $\eta_{11}$ (lower). Colors denote different values of the initial velocity $\dot{\gamma}$. The dashed line denotes the position of the respective spectral wall obtained from the spectral analysis. }
           \label{fig:trajs}
\end{figure}

\begin{figure}
    \centering
    \includegraphics[width=0.7\linewidth]{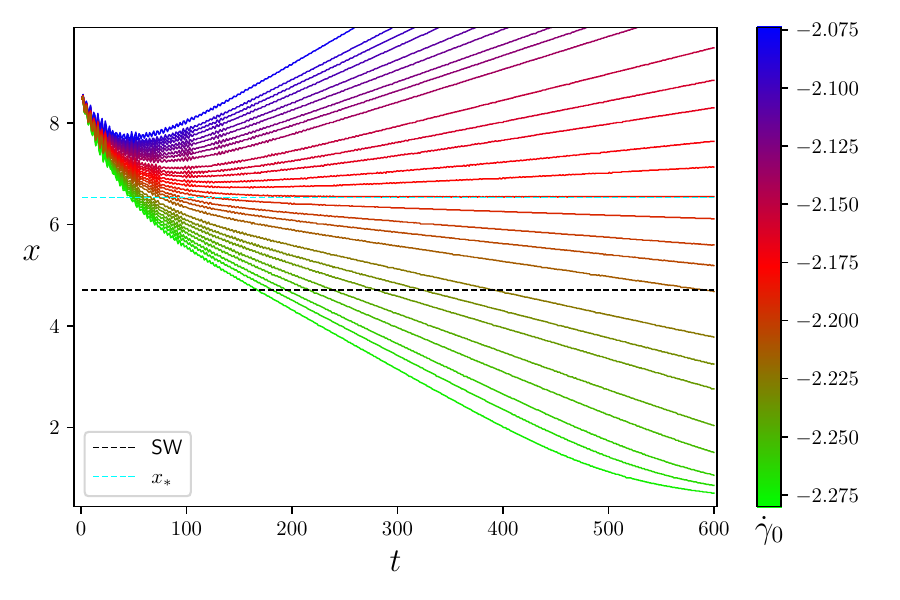}
    \caption{Trajectories for a sub-kink excited with the $n=9$ mode. The cyan line denotes the position of the static solution in the moduli space, the dashed black line denotes the postion of the spectral wall.}
    \label{fig:diss_crit}
\end{figure}

In Fig. \ref{fig:trajs} and \ref{fig:diss_crit}  we show the dependence of the position $x_0$ of one of the half-kinks in a molecule in the full field theory. We  start with the initial conditions \eqref{init_conds1} where 9th, 10th and 11th mode is respectively excited. We assume $\gamma_0=50$ (i.e., $x_0\approx 8.5$), $A_0=0.25$, and vary the initial velocity $\dot{\gamma_0}$. The position of the half-kink is defined as the point where $\phi=1/2$. We have performed the simulation in a grid of $1200$ points, from $x=-30$ to $x=30$ with $\delta x=0.05$.

First of all, it is clearly visible that the excitation of the mode leads to the appearance of the repulsive force. This effect happens for any excited mode. The sign of the force can be explained by the fact that the real frequency of the mode grows as the half-kinks, which form the solitonic BPS molecule, come closer to each other. For small velocities the force always overcomes the geodesic dynamics and the half-kinks are reflected back. The moment of the back scattering occurs sooner (for larger $x$) if the velocity is smaller. In other words, if the ratio between the kinetic and potential energy is smaller. For higher velocities (bigger ratio) the geodesic dynamics wins and the half kinks pass the $\gamma=0$ point. Note that the time where the force acts is $t \sim 100$. This agrees with our finding that $\Gamma \sim 10^{-2}$. Therefore the mode still possesses some energy and the force is strong enough to modify the geodesic dynamics. 

Secondly, we found that there is a very well visible spectral wall. For a critical value of the velocity $v_{cr}$ ($\dot{\gamma}_{cr}$) a quasi-stationary state is formed. It separates the back scattering motion from the passing through solutions. Its spatial location $x_0^{sw}$ agrees with a very good accuracy with the distance between the half-kinks (value of $\gamma$) at which the excited mode hits the mass threshold. Of course, this value is universal and does not dependent on the initial conditions. In Fig. \ref{fig:trajs} we excite $\eta_{10}$ and $\eta_{11}$ modes with initial frequencies $\omega_{10}^2=3.14$ and $\omega_{11}^2=3.49$. Positions of their spectral walls are respectively $x^{sw}_{10}=5.66$ and $x^{sw}_{11}=6.05$. Importantly, as it is known form previous works, the spectral wall is a very selective phenomenon. A BPS solution with an excited mode is affected only by the spectral wall corresponding to this particular mode and remains blind to all other spectral walls. 

The fact that we clearly see the spectral walls is undoubtedly connected with the previous observation of the decreasing of the decay constant as the mode approaches the mass threshold. In our simulations, the initial amplitude of the mode is rather big, as can be seen in the significant wobbling in the half-kink trajectories in the initial stage of the evolution. Then, as the soliton approaches the spectral wall, approximately at $t\sim 100$, the wobbling decreases substantially. Despite of that, the stationary solution can be formed and it lasts for quite a long time. Thus, the Feshbach resonance effectively behaves like a normal mode close to the spectral wall.

Finally, we identified a scenario whose existence is intimately related to the quasi-normal modes and has not been reported for solitonic dynamics with normal modes excited. During the collisions, the kinetic energy of the BPS solution may be transferred into the potential energy stored in the quasi-normal mode. This energy can be, however, radiated out due to exponential decay of the mode, leading to a BPS solution without both kinetic and internal energy. Hence, we find a static BPS solution as the final effect of the evolution. This effect is clearly visible in Fig. \ref{fig:diss_crit} where the 9th mode is excited. It seems that in order to have this phenomenon the decay constant should not be too small. On the contrary to the spectral wall, the position of this static solution is not universal but depends on the initial conditions i.e., the velocity and amplitude of the mode.

\section{Conclusions} 
\label{sec:conc}
In this work we have focused on a simple 1+1 dimensional model that presents a BPS sector of energetically degenerate soliton solutions, and whose spectrum of linearized perturbations presents quasinormal modes of the Feshbach type. We have studied the dynamics of such solitonic configurations, both from the numerical (through lattice field theory simulations) and semi-analytical (through the collective coordinate approximation) approaches, and found, as expected, that both methods are in agreement as long as none of the quasi-normal modes are excited. However, we have found that the excitation of these quasinormal modes, or Feshbach resonances, can modify the dynamics of solitons rather dramatically. In particular, we have found two major effects that these resonances have in the soliton dynamics: the first one consists of the standard mode-generated force due to the flow of the corresponding frequency eigenvalue along the moduli space. Such flow may have a great effect especially at the points in which the eigenfrequency meets the mass threshold, resulting in the formation of a long-living quasi-stationary state also known as spectral wall. Although the appearance of spectral walls and mode generated forces is quite established in soliton dynamics, we have for the first time confirmed that the same kind of effects can occur in the case of semi-bound excitations with an exponential decay. 

It is important to underline that the existence of the spectral wall is a consequence of both the flow of the mode eigenfrequency and the flow of its decay constant on the moduli space. The spectral wall requires that the frequency of a mode (in our case, the bounded part of the Feshbach resonance) hits the mass threshold. However, it is the flow of $\Gamma_n$, and, strictly speaking, the fact that it tends to 0 as the frequency approaches the mass threshold, which makes the existence of the long-lived quasi-stationary state possible. In other words,  the quasinormal modes become more and more stable as they approach their corresponding spectral walls.

The second important effect that we have observed is a combination between a resonant energy transfer between the kinetic degree of freedom of the soliton and the amplitude of the internal quasi-bound mode, and the fact that such modes decay much faster than standard bounded normal modes. The interplay of both effects results in a very efficient dissipation mechanism in which the solitons may lose most of their kinetic energy by transferring it to the amplitude of the rapidly-decaying resonance. As opposed to the spectral walls and mode-generated forces, the dissipative mechanism we have described can occur exclusively for models with solitons that present no normal modes, but semi-bound states instead, and should appear quite generically in this kind of models. Of course, further research is required to confirm this conjecture.

\vspace*{0.2cm}

The main aim of our analysis was to get some intuition for physically much more appealing situations, such as the dynamics of 't Hooft-Polyakov monopoles. Indeed, even a single monopole is known to carry infinitely many QNM. Therefore, two-monopole collisions should present similar phenomenology as that described in the present paper. Until now, the dynamics of BPS monopoles has been studied only in the unexcited regime which is fully covered by the geodesic dynamics. Although more complicated scenarios have recently been considered (see e.g., collisions of monopoles with domain walls \cite{Bachmaier:2023zmq,Bachmaier:2023wzz}), the monopoles have still been always assumed to be unexcited. 

There are no doubts that a mode generated force should exist for the excited BPS monopoles. However, in order to precisely predict its impact on the geodesic dynamics one has to know how the spectral structure changes as one flows on the moduli space. Therefore, the first step should be to study such a flow for the BPS two-monopole solution. Due to the integrability of the BPS regime, this may be even simpler than in the BPS 2-vortex case \cite{Alonso-Izquierdo:2023cua}. Nonetheless, taking into account the fact that the QNMs with real frequencies located close to the mass threshold of the vector field have very small decay constant we expect that the geodesic dynamics of multi-monopole system should be strongly affected by the excitation of such modes. We also expect that the spectral wall phenomenon will be present in collisions of BPS monopoles. 

\vspace*{0.2cm}

Lastly, the new dissipative mechanism may be of importance in the dynamics of solitons and defects in the cosmological context. 
An interesting idea to explore would be to consider whether the excitation of semi-bound states in such defects can trigger an energy loss mechanism which may alter the expected emission of radiation and gravitational waves from these objects.

\acknowledgments
JQ and AW has been supported in part by Spanish Ministerio de Ciencia e Innovación (MCIN) with
funding from the European Union NextGenerationEU (PRTRC17.I1) and the Consejería
de Educación, Junta de Castilla y León, through QCAYLE project, as well as the grant
PID2023-148409NB-I00 MTM, and the project Programa C2 from the University of Salamanca.
AGMC acknowledges support from the PID2021-123703NB-C21 grant funded by MCIN/AEI/10.13039/501100011033/and by ERDF;“ A way of making Europe”; and the Basque Government grant (IT-1628-22).

\bibliography{Bibliography.bib}

\begin{thebibliography}{10}

\bibitem{Manton:2004tk}
N.~S. Manton and P.~Sutcliffe.
\newblock {\em {Topological solitons}}.
\newblock Cambridge Monographs on Mathematical Physics. Cambridge University Press, 2004.

\bibitem{Samols:1991ne}
T.~M. Samols.
\newblock {Vortex scattering}.
\newblock {\em Commun. Math. Phys.}, 145:149--180, 1992.

\bibitem{tHooft:1974kcl}
Gerard 't~Hooft.
\newblock {Magnetic Monopoles in Unified Gauge Theories}.
\newblock {\em Nucl. Phys. B}, 79:276--284, 1974.

\bibitem{Polyakov:1974ek}
Alexander~M. Polyakov.
\newblock {Particle Spectrum in Quantum Field Theory}.
\newblock {\em JETP Lett.}, 20:194--195, 1974.

\bibitem{Manton:1981mp}
N.~S. Manton.
\newblock {A Remark on the Scattering of BPS Monopoles}.
\newblock {\em Phys. Lett. B}, 110:54--56, 1982.

\bibitem{Krusch:2024vuy}
Steffen Krusch, Morgan Rees, and Thomas Winyard.
\newblock {Scattering of vortices with excited normal modes}.
\newblock {\em Phys. Rev. D}, 110(5):056050, 2024.

\bibitem{AlonsoIzquierdo:CollectiveVortices}
A.~Alonso~Izquierdo, N.~S. Manton, J.~Mateos~Guilarte, and A.~Wereszczynski.
\newblock {Collective coordinate models for 2-vortex shape mode dynamics}.
\newblock {\em Phys. Rev. D}, 110(8):085006, 2024.

\bibitem{Campbell:1983xu}
David~K. Campbell, Jonathan~F. Schonfeld, and Charles~A. Wingate.
\newblock {Resonance structure in kink-antikink interactions in \ensuremath{\varphi} 4 theory }.
\newblock {\em Physica D}, 9:1, 1983.

\bibitem{Sugiyama:1979mi}
T.~Sugiyama.
\newblock {Kink-Antikink collisions in the two-dimensional phi**4 model}.
\newblock {\em Prog. Theor. Phys.}, 61:1550--1563, 1979.

\bibitem{Manton:2021ipk}
N.~S. Manton, K.~Oles, T.~Romanczukiewicz, and A.~Wereszczynski.
\newblock {Collective Coordinate Model of Kink-Antikink Collisions in \ensuremath{\phi}4 Theory}.
\newblock {\em Phys. Rev. Lett.}, 127(7):071601, 2021.

\bibitem{Alonso-Izquierdo:2023cua}
A.~Alonso-Izquierdo, W.~Garcia Fuertes, N.~S. Manton, and J.~Mateos~Guilarte.
\newblock {Spectral flow of vortex shape modes over the BPS 2-vortex moduli space}.
\newblock {\em JHEP}, 01:020, 2024.

\bibitem{Adam:2019xuc}
C.~Adam, K.~Oles, T.~Romanczukiewicz, and A.~Wereszczynski.
\newblock {Spectral Walls in Soliton Collisions}.
\newblock {\em Phys. Rev. Lett.}, 122(24):241601, 2019.

\bibitem{Alonso-Izquierdo:SWvortices}
A.~Alonso-Izquierdo, J.~Mateos~Guillarte, M.~Rees, and A.~Wereszczynski.
\newblock {Spectral wall in collisions of excited Abelian Higgs vortices}.
\newblock {\em Phys. Rev. D}, 110(6):065004, 2024.

\bibitem{Forgacs:2003yh}
Peter Forgacs and Mikhail~S. Volkov.
\newblock {Resonant excitations of the 't Hooft-Polyakov monopole}.
\newblock {\em Phys. Rev. Lett.}, 92:151802, 2004.

\bibitem{Russell:2010xx}
Katie~M Russell and Bernd~J Schroers.
\newblock {On resonances and bound states of the 't Hooft-Polyakov monopole}.
\newblock {\em Phys. Rev. D}, 83:065004, 2011.

\bibitem{Feshbach}
Herman {Feshbach}.
\newblock {Unified theory of nuclear reactions}.
\newblock {\em Annals of Physics}, 5(4):357--390, December 1958.

\bibitem{PhysRev.147.73}
Marvin~H. Mittleman.
\newblock Resonances in multichannel scattering.
\newblock {\em Phys. Rev.}, 147:73--76, Jul 1966.

\bibitem{Alonso-Izquierdo:2024tjc}
A.~Alonso-Izquierdo, J.~J. Blanco-Pillado, D.~Migu\'elez-Caballero, S.~Navarro-Obreg\'on, and J.~Queiruga.
\newblock {Excited Abelian-Higgs vortices: Decay rate and radiation emission}.
\newblock {\em Phys. Rev. D}, 110(6):065009, 2024.

\bibitem{Berti:2009kk}
Emanuele Berti, Vitor Cardoso, and Andrei~O. Starinets.
\newblock {Quasinormal modes of black holes and black branes}.
\newblock {\em Class. Quant. Grav.}, 26:163001, 2009.

\bibitem{Dorey:2017dsn}
Patrick Dorey and Tomasz Roma\'nczukiewicz.
\newblock {Resonant kink-antikink scattering through quasinormal modes}.
\newblock {\em Phys. Lett. B}, 779:117--123, 2018.

\bibitem{Campos:2019vzf}
Jo\~ao G.~F. Campos and Azadeh Mohammadi.
\newblock {Quasinormal modes in kink excitations and kink\textendash{}antikink interactions: a toy model}.
\newblock {\em Eur. Phys. J. C}, 80(5):352, 2020.

\bibitem{Bazeia:1995en}
D.~Bazeia, M.~J. dos Santos, and R.~F. Ribeiro.
\newblock {Solitons in systems of coupled scalar fields}.
\newblock {\em Phys. Lett. A}, 208:84--88, 1995.

\bibitem{Bazeia:1996np}
D.~Bazeia and M.~M. Santos.
\newblock {Classical stability of solitons in systems of coupled scalar fields}.
\newblock {\em Phys. Lett. A}, 217:28--30, 1996.

\bibitem{ShifmanVoloshin}
M.~A. Shifman and M.~B. Voloshin.
\newblock Degenerate domain wall solutions in supersymmetric theories.
\newblock {\em Phys. Rev. D}, 57:2590--2598, Feb 1998.

\bibitem{Alonso-Izquierdo:2013isa}
A.~Alonso-Izquierdo, D.~Bazeia, L.~Losano, and J.~Mateos~Guilarte.
\newblock {New Models for Two Real Scalar Fields and Their Kink-Like Solutions}.
\newblock {\em Adv. High Energy Phys.}, 2013:183295, 2013.

\bibitem{phi6scattering}
C.~Adam, P.~Dorey, A.~Garcia Martin-Caro, M.~Huidobro, K.~Oles, T.~Romanczukiewicz, Y.~Shnir, and A.~Wereszczynski.
\newblock {Multikink scattering in the \ensuremath{\phi}6 model revisited}.
\newblock {\em Phys. Rev. D}, 106(12):125003, 2022.

\bibitem{Manton:1996ex}
N.~S. Manton and H.~Merabet.
\newblock {Phi**4 kinks: Gradient flow and dynamics}.
\newblock {\em Nonlinearity}, 10:3, 1997.

\bibitem{Blanco-Pillado:2021jad}
Jose~J. Blanco-Pillado, Daniel Jim\'enez-Aguilar, Jose~M. Queiruga, and Jon Urrestilla.
\newblock {Internal excitations of global vortices}.
\newblock {\em JCAP}, 10:047, 2021.

\bibitem{Bachmaier:2023zmq}
Maximilian Bachmaier, Gia Dvali, and Juan~Sebasti\'an Valbuena-Berm\'udez.
\newblock {Radiation emission during the erasure of magnetic monopoles}.
\newblock {\em Phys. Rev. D}, 108(10):103501, 2023.

\bibitem{Bachmaier:2023wzz}
Maximilian Bachmaier, Gia Dvali, Juan~Sebasti\'an Valbuena-Berm\'udez, and Michael Zantedeschi.
\newblock {Confinement slingshot and gravitational waves}.
\newblock {\em Phys. Rev. D}, 110(1):016001, 2024.

\end{thebibliography}

\end{document}